\documentclass[aps, floatfix,nofootinbib,amsmath,amssymb,amsfonts,notitlepage, twocolumn,superscriptaddress,prl]{revtex4-1} 
\usepackage[english]{babel}
\usepackage[utf8]{inputenc}
\usepackage[colorinlistoftodos, color=green!40, prependcaption]{todonotes}

\begin{document}
\title{Observation of non-Markovian Radiative Phenomena in Structured Photonic Lattices}

\author{Rodrigo A. Vicencio}
    \email{rvicencio@uchile.cl}
    \affiliation{Departamento de Física, Facultad de Ciencias Físicas y Matemáticas, Universidad de Chile, Santiago, Chile}
    \affiliation{Millennium Institute for Research in Optics - MIRO, Santiago, Chile}

\author{Fabiola G.L. C\'arcamo-Macaya}
    \affiliation{Departamento de Física, Facultad de Ciencias Físicas y Matemáticas, Universidad de Chile, Santiago, Chile}

    \author{Diego Rom\'an-Cort\'es}
    \affiliation{Departamento de Física, Facultad de Ciencias Físicas y Matemáticas, Universidad de Chile, Santiago, Chile}
    \affiliation{Millennium Institute for Research in Optics - MIRO, Santiago, Chile}

    \author{Pablo Solano}
\email{psolano@udec.cl}
\affiliation{Departamento de F\'isica, Facultad de Ciencias F\'isicas y Matem\'aticas, Universidad de Concepci\'on, Concepci\'on, Chile}

\date{\today} 

\begin{abstract}
The spectral structure of a photonic reservoir shapes radiation phenomena for embedded quantum emitters. We implement an all-optical analogue to study such an effect, particularly to observe the non-Markovian radiation dynamics of an emitter coupled to two-dimensional structured reservoirs. Its dynamics is simulated by light propagating through a photonic lattice, acting as a reservoir for an adjacent waveguide that mimics a coupled quantum emitter. We study radiation dynamics in square and Lieb lattices under different coupling regimes and observe how the flat band properties of the Lieb lattice significantly enhances light-matter coupling and non-Markovianity. Our platform opens a path for the experimental exploration of single photon quantum optical phenomena in structured reservoirs to enhance light-matter interactions.
\end{abstract}


\maketitle

The interaction between quantum emitters (QEs) and photonic modes plays a central role in quantum optics foundations and applications. Tailoring such interaction, by modifying the dispersion relation and dimensionality of the photonic medium~\cite{John1990,John1994,Lodahl2015,Wu2018,Chang2018}, can lead to enhanced~\cite{John1995}, suppressed~\cite{Angelakis2004}, and directional interactions~\cite{Lodahl2017}, and non-trivial radiation dynamics~\cite{Vats1998,Lambropoulos2000,Gonzales-Tudela2017,Gonzales-Tudela2017b}. The band structure of the reservoir determines the density of states available to a coupled QE, and hence its interaction strength. For example, placing a QE on resonance with the band edge of a photonics reservoir enhances light-matter interactions~\cite{Bello2019,Ferreira2021}. More recently, reservoirs with flat bands (FBs) that have, in general, a set of dispersive and non-dispersive bands~\cite{Flachreport,FBreport} with extended and compact states~\cite{vicencio_observation_2015,Mukherjee2015} respectively, have emerge as candidates to enhance such coupling~\cite{DeBernardis2021,Bienias2022,DeBernardis2023,Dibenedetto2024}. 

Significant coupling and energy exchange between the QE and the reservoir manifest as non-Markovian radiative decay, a central example of the rich phenomenology lying beyond traditional quantum optics enabled by structured reservoirs~\cite{deVega2017,Quang1997,John1997,John2001,Gonzales-Tudela2017,Sinha2020}. While increasingly promising, embedding QEs in 2D materials with non trivial band structures faces practical challenges that complicate experimental explorations of novel reservoir structures and collective effects of many indistinguishable emitters~\cite{Azzam2021}. 

\begin{figure}[t]
    \centering
    \includegraphics[width=0.95\linewidth]{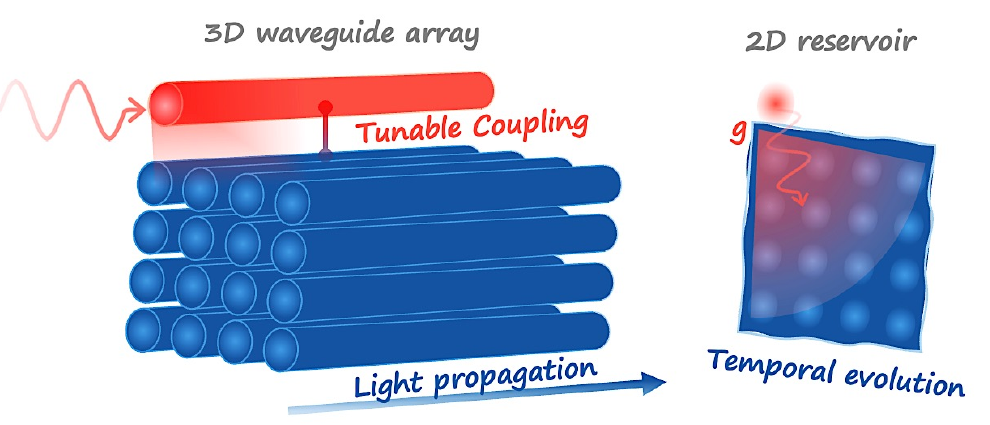}
    \caption{Light propagating through a three-dimensional waveguide array simulates the temporal evolution of a quantum emitter coupled to a two-dimensional structured reservoir. Tunable couplings allow the experimental exploration of a wide range of coupling regimes.}
    \label{fig:1}
\end{figure}

Photonic waveguide arrays (PWA) offer a versatile platform to simulate QEs coupled to structured photonic reservoirs~\cite{Lambropoulos2000,Longhi2009,Crespi2012,Crespi2019,kagome2024}. In such platforms, an array of evanescently coupled waveguides simulates a reservoir, while a nearby individual waveguide plays the role of a QE, which decays into the reservoir (see Fig.~\ref{fig:1}). Light propagation through the whole PWA acts as an optical analogue for the temporal evolution of the radiation phenomena~\cite{Crespi2019}. In this sense, a PWA can reproduce quantum optical effects from cavity QED to waveguide QED in the weak and strong coupling regimes, in the same manner that coupled cavities arrays~\cite{Calajo2016}.  Although PWA cannot simulate quantum systems in the nonlinear multiphoton regime, they can simulate collective effects of indistinguishable emitters~\cite{kagome2024} and the radiation dynamics of delayed-induced non-Markovian reservoirs~\cite{Longhi2020a}. Nonetheless, PWAs have an untapped potential to study the radiation phenomena of QEs coupled to arbitrary structured photonics reservoirs. 

In this paper, we experimentally study the decay dynamics of a single quantum emitter coupled to two-dimensional structured reservoirs, in an optical analogue system. We explore square and Lieb lattice reservoirs, where a nearby evanescently coupled waveguide excited by a laser plays the role of the quantum emitter. The light propagating through the array simulates the temporal evolution of the system, evidencing non-Markovianity through the re-excitation of the radiating emitter (waveguide). The variation of the distance in between the emitter and the reservoir allows us to observe dynamical behaviors corresponding to those that characterize the weak and strong coupling regimes. We observe how the non-Markovian decay predicted for a square lattice~\cite{Gonzales-Tudela2017}, whose linear spectrum is  dispersive and manifests only transport~\cite{lederer_discrete_2008}, becomes more significant for a zero-dispersion flat-band, such as the one of a Lieb lattice~\cite{vicencio_observation_2015,Mukherjee2015}. Our results present photonic waveguide arrays as a versatile platform to study single photon quantum optical phenomena of theoretical and practical relevance for quantum optics and quantum information processing.


As the Fermi's Golden Rule (FGR) describes, the radiation dynamics of a QE depends on the density of states (DOS) of a given reservoir. However, the FGR relies on a perturbative description of the dynamics in the Markov approximation. When the DOS diverges, naively considering the FGR would lead to a nonphysical, infinitely fast decay rate. Beyond the perturbative regime, a diverging DOS leads to an effectively stronger coupling to the reservoir, and the system becomes non-Markovian. A signature of this non-Markovianity is a non-exponential decay and oscillations of the QE excitation probability.  

The physics described above can be realized in an analogue optical system, using PWA with diverging DOS. Such arrays can simulate a 2D structured reservoir, while an extra waveguide can mimic a QE coupled to it. Lets us consider a waveguide array described by the Hamiltonian $\hat{H}_{\text{B}}=\hbar\omega_{\text{B}}\sum_{\bf{n}} \hat{b}^{\dagger}_{\bf{n}} \hat{b}_{\bf{n}}+V_1\sum_{\langle \bf{n},\bf{m}\rangle}\left(\hat{b}^{\dagger}_{\bf{n}} \hat{b}_{\bf{m}}+\hat{b}^{\dagger}_{\bf{m}} \hat{b}_{\bf{n}}\right)$, where $\hat{b}_{\bf{n}}$ ($ \hat{b}^{\dagger}_{\bf{m}}$) is the bosonic bath operator of the $\bf{n}$($\bf{m}$)-th waveguide, $\hbar\omega_{\text{B}}$ is the on-site lattice energy, and $V_1$ is the coupling constant in between adjacent waveguides. The QE is simulated by a single nearby waveguide with Hamiltonian $\hat{H}_{\text{QE}}=\hbar\omega_{0} \hat{b}^{\dagger}_{\bf{0}}\hat{b}_{\bf{0}}$. Such waveguide is evanescently coupled to the lattice with a coupling constant $V_2$ by the interaction Hamiltonian $\hat{H}_{\text{int}}=V_2\left(\hat{b}^{\dagger}_{\bf{0}} \hat{b}_{\bf{n}_0}+\hat{b}^{\dagger}_{\bf{n}_0} \hat{b}_{\bf{0}}\right)$, with $\bf{n}_0$ the lattice site coupled to the analog QE. The bath Hamiltonian can be diagonalized in $\textbf{k}$ space, considering $\hat{b}_n=\frac{1}{\sqrt{N}}\sum_{\textbf{k}}\hat{a}_{\textbf{k}}e^{i \textbf{k}\cdot\textbf{n}}$, obtaining
\begin{eqnarray}
\hat{H}_{\text{B}}&=&\sum_{\textbf{k}}\hbar\omega(\textbf{k}) \hat{a}^{\dagger}_{\textbf{k}}\hat{a}_{\textbf{k}}\ , \quad \hat{H}_{\text{QE}}=\hbar\omega_{0} \hat{b}^{\dagger}_{\bf{0}} \hat{b}_{\bf{0}}\ , \nonumber \\ \hat{H}_{\text{int}}&=&\sum_{\textbf{k}}\frac{V_2}{\sqrt{N}}\left( \hat{a}_{\textbf{k}}\hat{b}^{\dagger}_{\bf{0}}+\hat{a}^{\dagger}_{\textbf{k}} \hat{b}_{\bf{0}}\right)\ ,
\label{eq: PWA H}
\end{eqnarray}
where $\omega(\textbf{k})$ is the dispersion relation (band) of the lattice. 

\begin{figure}[t!]
    \centering
    \includegraphics[width=1\linewidth]{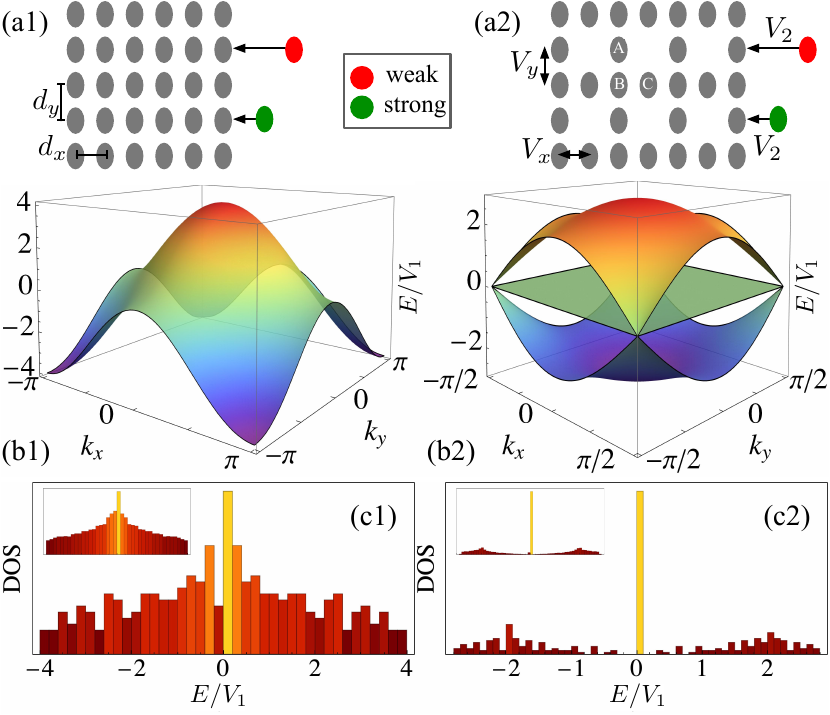}\vspace{-0.5cm }
    \caption{(a1) Square and (a2) Lieb lattice geometries, including a sketch for weak (red) and strong (green) coupling regimes. (b) Band spectra for (b1) Square and (b2) Lieb lattices, for $V_x=V_y=V_1=1$ cm$^{-1}$. (c) DOS histograms for (c1) square and (c2) Lieb lattices, for $361$ and $280$ lattice sites ($2601$ and $1976$ sites at insets), respectively.}
    \label{fig:2}
\end{figure}

These equations are equivalent to those describing a QE coupled to a reservoir \cite{Lambropoulos2000,Gonzales-Tudela2017,Crespi2019,Calajo2016} in the single excitation regime, where the QE spin operator $\hat{\sigma}_{-(+)}$ is equivalent to the bosonic operator $\hat{b}^{(\dagger)}_{0}$\cite{SM}. The main difference is that, instead of describing the temporal evolution with the Schrödinger's equation, model~(\ref{eq: PWA H}) describes light propagation along the $z$ coordinate of a PWA (see Fig.~\ref{fig:1}) via a Discrete Linear Schrödinger Equation (DLSE) \cite{Lederer2008}. As photons do not interact on a linear level, injecting an intense light beam trough the PWA and taking an output image at a propagation distance $z$ is equivalent to measuring multiple photo-detections, spontaneously emitted by the QE, after a fixed time $t$. Hence, the decay dynamics of a single QE coupled to a structured reservoir can be reconstructed by collecting the images at the output of the PWA at different propagation distances.

We consider two different PWA to simulate 2D structured reservoirs with a diverging DOS: square and Lieb lattices, as shown in Fig.~\ref{fig:2}(a). The dispersion relation of a square lattice is formed by a single band given by $E(k_x,k_y)=\hbar\omega(k_x,k_y)=2V_x \cos (k_x d_x) +2V_y \cos (k_y d_y)$, with $V_x$ and $V_y$ the horizontal and vertical coupling coefficients depending on distances $d_x$ and $d_y$ [see Fig.~\ref{fig:2}(a)]. This spectrum [shown in Fig.~\ref{fig:2}(b1)] is characterized by a saddle point and large degeneracy at zero energy, corresponding to different combinations of $k_x$ and $k_y$ wave-vectors. On the other hand, a Lieb lattice has three sites per unit cell [see sites $A$, $B$ and $C$ in Fig.~\ref{fig:2}(a2)], with a dispersion relation given by $E(k_x,k_y)=0,\ \pm 2\sqrt{V_x^2 \cos^2 (k_x d_x) +V_y^2 \cos^2 (k_y d_y)}$; i.e., a spectra composed of two dispersive bands and one FB at zero energy~\cite{vicencio_observation_2015}. This band structure has a strong degeneracy for all wavenumbers at $E=0$ [see Fig.~\ref{fig:2}(b2)], with a number of FB modes corresponding to $\sim 1/3$ of the spectrum. However, realistic lattice implementations are finite and have a finite DOS. Fig.~\ref{fig:2}(c) shows the DOS (energy histogram) for both lattices, both with peaks at $E=0$, for a lattice area of $19\times19$ sites. We observe DOS having clear distributions, which are of course better defined when increasing the lattice dimensions [see insets in Fig.~\ref{fig:2}(c)]. A square lattice has a large accumulation of states around $E=0$, with a smoothly decaying distribution up to the band edges located at $E=\pm4V_1$. On the other hand, a Lieb lattice has a very contrasting DOS, with a huge peak (divergence) exactly at $E=0$, and two symmetric distributions at around $E=\pm2V_1$.

It can be insightful to describe the origin of non-Markovian dynamics in terms of the group velocity $v_g(\textbf{k})= \nabla E(\textbf{k})$, which describes how fast the emitted excitation could propagate through the reservoir. Since $v_g$ is inversely proportional to the DOS, one can interpret the non-Markovian effects of a large DOS in terms of reduced $v_g$. If the $v_g$ is larger (smaller) than the coupling rate of the QE, $V_2$, then the QE cannot (can) be re-excited from the reservoir, leading to (non-)Markovian dynamics. Such dynamics characterizes the weak and strong coupling regimes~\cite{Calajo2016}. Strong coupling shows Rabi-like oscillations of the QE while decaying into the lattice reservoir, while weak coupling shows an (almost) exponential decay. Since in the square lattice $v_g=2V_1$ at $E=0$, we will simply observe the weak and the strong coupling regimes for $V_2\lesssim V_1$ and $V_2\gtrsim V_1$, respectively. On the other hand, in the case of a Lieb lattice, the FB automatically provides a zero group velocity and, therefore, an enhanced non-Markovian behavior, regardless of how small $V_2$ is. For the ideal case of a QE coupled only to a Lieb FB ($\omega(\textbf{k})=\omega_{FB}=0$), we expect a simple oscillatory dynamics for the amplitude of the QE~\cite{SM}
 \begin{equation}
c_{QE}(z)=\cos[V_2 z/(2\sqrt{N})].
\end{equation}
However, on a real system, we have to also include the excitation of the dispersive bands, and the radiative process will be indeed a result of a mixed of decaying and oscillatory dynamics.

\begin{figure}[t!]
    \centering
    \includegraphics[width=1\linewidth]{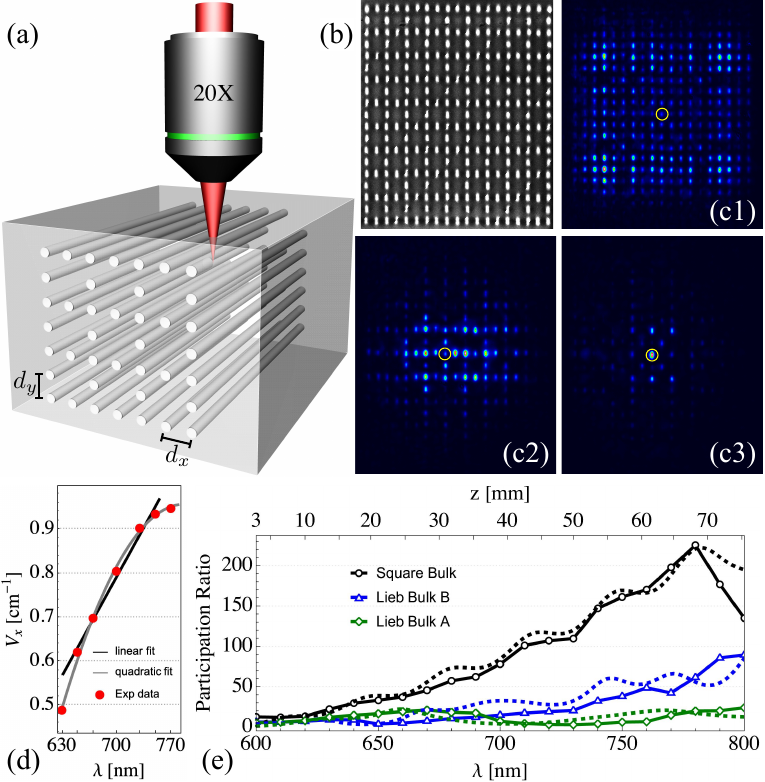}
    \caption{(a) Femtosecond writing technique and (b) white light microscope image of a Lieb lattice. (c) Output intensity profiles after a bulk excitation: (c1) square lattice, (c2) $B$ and (c3) $A$ sites of a Lieb lattice, at  $\lambda = 730$ nm. (d) $V_x$ at $d_x=17\ \mu$m vs $\lambda$. (e) Experimental (symbols connected by lines) and numerical (dashed lines) participation ratio evolution versus wavelength (bottom axis) and propagation distance (top axis) for the excitations described in (c).}
    \label{fig:3}
\end{figure}


We fabricate different PWA using the femtosecond laser-writing technique~\cite{szameit_discrete_2005}, on a $L = 5$ cm long borosilicate glass wafer, as sketched in Fig.~\ref{fig:3}(a). Fig.~\ref{fig:3}(b) shows an example of a fabricated Lieb lattice with 280 single-mode waveguides and inter-site distances $\{d_x,d_y\}=\{17,18.5\}\ \mu$m, such that $V_x/ V_y\approx 1$~\cite{SM}. Fig.~\ref{fig:3}(c1) shows a symmetric diffraction pattern for an excitation wavelength $\lambda = 730$ nm, after exciting a central bulk site of a square lattice. However, the output pattern of a Lieb system strongly depends on the input site~\cite{NJP_2014}, as the FB states occupy sites A and C only. Therefore, a bulk $B$ ($C$) excitation will show mostly transport (localization), as shown in Figs.~\ref{fig:3}(c2) and (c3). 

We characterize the coupling constants experimentally~\cite{SM}, by fabricating several couplers with different inter-site distances, and by exciting them with different wavelengths by means of a supercontinuum (SC) laser source. Fig.~\ref{fig:3}(d) shows a monotonous increasing tendency of $V_x$ versus $\lambda$~\cite{Szameit07,Szameit16,diamondAPL23,prb24,grafenoDipolar,SM}, which is very well described by a quadratic function $V_x(\lambda)=a+b\lambda+c\lambda^2$ (see gray line). The optimal performance of our PWA is found for $\lambda\in\{650,750\}$ nm, where the guiding properties of our waveguides are enhanced. In this range, we can approximate $V_x(\lambda)\sim \lambda$ [black line in Fig.~\ref{fig:3}(d)]. Now, we characterize the bulk diffraction properties of the fabricated lattices using a wavelength scan method~\cite{noh_topological_2018,diamondAPL23,prb24,grafenoDipolar,SM}, in the wavelength range $\lambda\in\{600,800\}$ nm, and for a fixed propagation length $L$. The data extracted directly from the experiments~\cite{SM} is presented in Fig.~\ref{fig:3}(e) by symbols connected with full lines. There, we show the evolution over the excitation wavelength of the participation ratio defined as $R\equiv (\sum_{\bf{n}} P_{\bf{n}})^2/\sum_{\bf{n}} P_{\bf{n}}^2$, with $P_{\bf{n}}=|u_{\bf{n}}|^2$ the light intensity at the $\bf{n}$-th site ($R$ measures the number of effectively excited sites of a given array). The data was obtained for a single-site bulk excitation of a square lattice (black), and bulk $B$ (blue) and bulk $C$ (green) sites of a Lieb system. We observe how a square lattice manifests its transport-only properties with a strong and fast dispersion, and a clear spreading through the whole array (large $R$ values). On the other hand, a Lieb lattice disperses more slowly, with the energy spreading over a comparative smaller region for a bulk $B$ excitation. For a bulk $A$ input condition the light keeps oscillating on a reduced spatial area only, manifesting mostly localization due to a more efficient excitation of FB states.

We numerically solve model (\ref{eq: PWA H}) over the propagation coordinate $z$, for $V_1=V_x=V_y=0.9$ cm$^{-1}$, obtained at $\lambda=730$ nm~\cite{SM}. We compare the numerical and experimental evolutions directly in Fig.~\ref{fig:3}(e). For the three different bulk excitations, we obtain an excellent agreement after using the linear transformation $\lambda=\lambda_0+\alpha z$~\cite{SM}. For this, we simply adjusted the data obtained from the discrete simulations to match the glass length $L=5$ cm for the coupling constant obtained at $\lambda=730$ nm. Therefore, a sweep in $\lambda$ can be interpreted as a linear variation of an effective normalized propagation distance $V_1 L$, and we can directly assume that the system size increases for a larger excitation wavelength. We corroborate this by performing intensive continuous numerical simulations on a square geometry and obtaining the light propagation along $z$ for different wavelengths~\cite{SM}. We find a clear monotonous increment of $R$ for a larger distance or larger wavelength, with a linear correspondence among those quantities, and an almost equal dispersion in the horizontal and vertical directions. Both, the discrete and continuous, results give a strong support for a direct analogy in between $z$ and $\lambda$ for 2D lattice systems.

\begin{figure}[]
    \centering
    \includegraphics[width=1\linewidth]{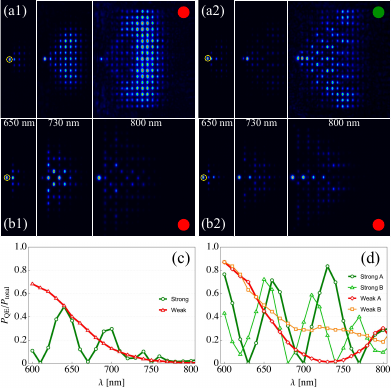}
    \caption{Output intensity profiles for a square lattice under (a1) Weak and (a2) Strong coupling conditions, and for a Lieb lattice under Weak coupling for (b1) $n_0=A$ and (b2) $n_0=B$ excitations, for the indicated excitation wavelengths. (c) and (d) Fraction of the power remaining at the QE versus the excitation wavelength for square and Lieb lattices, respectively.}
    \label{fig:4}
\end{figure}

We emulate the radiation phenomena of a single QE coupled to a 2D structured reservoir by evanescently coupling a single waveguide to the PWA, as it is sketched in Fig.~\ref{fig:2}(a). The distance in between this QE-waveguide and the PWA defines the strength of the coupling $V_2$: when the QE-waveguide is closer (further) to the PWA, the system is in the strong (weak) coupling regime (experimentally speaking, $V_2/V_1\approx1.5$ ($0.5$)~\cite{SM}). In the weak coupling regime, the square lattice [see Figs.~\ref{fig:4}(a1) and (c)-red] decays quasi-exponentially, and the excitation initially placed at the QE-waveguide spreads through the entire reservoir, showing a Markovian behavior. In contrast, in the strong coupling regime, a part of the excitation returns from the reservoir into the QE-waveguide, as it is shown in Figs.~\ref{fig:4}(a2) and (c)-green. The oscillation of the QE excitation probability is a signature of non-Markovianity, where the current state of the QE depends on its previous state.

The dynamics of radiation phenomena on a Lieb lattice is contrastingly different. For equal QE-reservoir coupling strength $V_2$, the Lieb lattice shows striking non-Markovian behavior originated from its FB. A QE-waveguide weakly coupled to a Lieb-reservoir, through an $A$ site, experiences a resonant interaction between the QE and the compact Lieb state~\cite{vicencio_observation_2015,Mukherjee2015} laying just beside. Both, the QE and the Lieb mode, have an energy $E$ equal to zero and are, therefore, perfectly tuned. Figs.~\ref{fig:4}(b1) and ~\ref{fig:4}(d)-red show that when the light tries to radiate into the reservoir it mostly excites the Lieb mode, at it is observed at $\lambda\sim 730$ nm. This mode then excites back the QE, which is again strongly populated at $\lambda\sim 800$ nm. This process is not isolated and has an effective loss due to the dispersive part of the spectrum excited, with light also radiating (diffracting) away through the lattice. On the other hand, the radiation of a weakly coupled QE through a $B$ site shows a similar dynamics, but with a very slow dissemination of the energy from the QE [see Figs.~\ref{fig:4}(b2) and ~\ref{fig:4}(d)-orange]. In this case, there is an excitation of an impurity-like state, having also a zero energy, which decays rather linearly from the QE into the lattice. As this state also excites the FB modes via the excitation of $C$ sites, the localization tendency inhibits much transport to the rest of the reservoir and the power ratio at the QE remains rather stacked from $\lambda\sim 700$ nm. Figure~\ref{fig:4}(d) also shows the strong coupling regime (green data). In this case, we observe well defined Rabi-like oscillations, with only few losses for $n_0=A$, originated from the excitation of impurity-like states emerging at the emitter interacting region~\cite{SM,prb24}. The reduced amplitude in the the cosine square evolution of the QE reflects that a fraction of the excitation can couple to the dispersive bands of the Lieb lattice. All the scenarios show clear evidence of non-Markovian radiation dynamics on a FB Lieb system, independent of the coupling regime.


In conclusion, we have implemented a photonic waveguide array to simulate strong non-Markovian effects in the decay dynamics of a quantum emitter coupled to 2D structured reservoirs. Our results show the different behaviors of an emitter decaying into a square and Lieb lattice reservoirs, in the weak and strong coupling regimes, highlighting the significant non-Markovianity produced by the enhanced light-matter coupling provided by the flat band of the Lieb lattice. We explicitly demonstrate an oscillatory dependence on the decaying behavior of a quantum emitter coupled to a FB and corroborate this experimentally. We probe the validity of a rather new experimental technique in 2D lattices, based on a wavelength scan method as an effective analog of the propagation coordinate. Beyond the particular observations in this work, we combine two fields of research, i.e., quantum optics and photonic lattices, to present a versatile platform for simulating quantum optical phenomena in structured reservoirs. Although the system is limited to exploring quantum optics in the single particle regime, it offers versatility in exploring various reservoirs and coupling cases. Furthermore, it allows the simulation of coherent quantum phenomena with analogs to many perfectly indistinguishable quantum emitters coupled to a common reservoir. We hope this technique could facilitate the research of novel collective quantum optical phenomena in structured reservoir beyond the FB case, including for example topological and non-Hermitian phenomena. 


\textit{Acknowledgements}.---
Authors acknowledge Kanu Sinha for stimulating discussions. This work was supported in parts by Millennium Science Initiative Program ICN17$\_$012 and ANID FONDECYT Grants 1231313 and 1240204.


%

\end{document}